\documentclass[aps,prl,twocolumn,groupedaddress,nofootinbib]{revtex4-1}

\usepackage{graphicx}
\usepackage{mathrsfs}

\newcommand{\scri}{\mathscr{I}}
\usepackage{color}

\begin{document}

\title{Ultimate Black Hole Recoil: What is the maximum high energy collisions kick?}

\author{James Healy}
\author{Carlos O. Lousto}

\affiliation{Center for Computational Relativity and Gravitation (CCRG),
School of Mathematical Sciences,
Rochester Institute of Technology, 85 Lomb Memorial Drive, Rochester,
New York 14623}

\date{\today}

\begin{abstract}
We performed a series of 1381 full numerical simulations of high
energy collision of black holes to search for the maximum recoil
velocity after their merger.  We consider equal mass binaries with
opposite spins pointing along their orbital plane and perform a search
of spin orientations, impact parameters, and initial linear momenta to
find the maximum recoil for a given spin magnitude $s$. This spin
sequence for $s=0.4, 0.7, 0.8, 0.85, 0.9$ is then extrapolated to the
extreme case, $s=1$, to obtain an estimated maximum recoil velocity of
$28,562\pm 342$ km/s, thus approximately bounded by $10\%$ the speed
of light.
\end{abstract}

\maketitle


\section{Introduction\label{sec:introduction}}

Ever since the discovery through full numerical simulations
\cite{Campanelli:2007ew,Gonzalez:2007hi} that the merger of binary
black holes may lead to large (astrophysically speaking) gravitational
recoil velocities, a fascinating search for such events in nature
takes place~\cite{Komossa:2012cy,Chiaberge:2018lkg}.  Since the first
modeling of large recoils~\cite{Campanelli:2007cga}, it was clear that
the spins of the black holes played a crucial role in their merger
remnant reaching up to several thousand km/s speeds. It was next found
a configuration \cite{Lousto:2011kp} that maximized the recoil nearing
5,000 km/s. This configuration combined the opposite spins of
\cite{Campanelli:2007cga} that maximized asymmetry with the hangup
effect \cite{Campanelli:2006uy} that maximized radiation.  All those
configurations assumed negligible eccentricities at the time of
merger, when most of the asymmetric radiation takes place. While this
is the most plausible astrophysical scenario, new gravitational waves
observations show the potential for large residual eccentricity in
some events \cite{Gayathri:2020coq}.

Here we will explore the extreme scenario of high energy collisions of
black holes, in the realm of high-energy colliders, to discover the
fundamental laws of nature \cite{Cardoso:2012qm,Berti:2016rij}, with
applications to the gauge/gravity duality, holography
\cite{Cardoso:2014uka}, primordial black hole collisions in the early
universe \cite{Franciolini:2022htd,Ding:2020ykt,Cai:2019igo}, and as
tests of the radiation bounds theorems and cosmic censorship
conjecture in General Relativity
\cite{Hawking:1971tu,Eardley:2002re,Siino:2009vw}.  The growth of
structure seeded by primordial black holes has been studied in
\cite{Mack:2006gz}, and the effects of gravitational-wave recoil on
the dynamics and growth of supermassive black holes has been studied
in \cite{Blecha:2008mg}. While the scenario of supermassive rotating
black holes potentially accelerating orbiting black holes to high
energies was discussed in \cite{Harada:2014vka}.

This high energy collision of black holes scenario was studied in
\cite{Sperhake:2008ga} to compute the maximum energy radiated by equal
mass, nonspinning black holes in an ultrarelativistic headon
collision. This first study was then followed up by the claim that the
spin effects did not matter for these collisions in
\cite{Sperhake:2012me}.  In \cite{Healy:2015mla} we then revisited the
headon scenario using new initial data \cite{Ruchlin:2014zva} with
low spurious initial radiation content that allows for more accurate
estimates of the maximum energy radiated placing it at about $13\%$.
Non headon high energy collisions have also been studied in
\cite{Shibata:2008rq}, and in notable analytic detail in
\cite{Berti:2010ce}.  Some of the early reviews on the subject are
\cite{Cardoso:2012qm,Berti:2016rij}, and more up-to-date ones are
\cite{Sperhake:2019oaw,Bozzola:2022uqu,Page:2022bem}.

Here we extend those studies with a much larger numerical simulations set
by directly solving numerically the General Relativity field equations in
supercomputers, and by focusing
on the computation of the maximum achievable gravitational recoil from
grazing, high energy collisions of binary black holes, where the
holes' spin orientation and magnitude play a crucial role.

\section{Numerical Techniques\label{sec:NR}}

The full numerical simulations were performed using the {\sc LazEv}
code~\cite{Zlochower:2005bj} implementation of the moving puncture
approach~\cite{Campanelli:2005dd}. We use the general relativistic
BSSNOK formalism of evolutions systems~\cite{Nakamura87, Shibata95,
  Baumgarte99}.  The {\sc LazEv} code uses the {\sc
  Cactus}~\cite{cactus_web} / {\sc Carpet}~\cite{Schnetter-etal-03b} /
{\sc EinsteinToolkit}~\cite{Loffler:2011ay, einsteintoolkit}
infrastructure.  The {\sc Carpet} mesh refinement driver provides a
``moving boxes'' style of mesh refinement.  To compute the numerical
(Bowen-York) initial data, we use the {\sc
  TwoPunctures}~\cite{Ansorg:2004ds} code.  We use {\sc
  AHFinderDirect}~\cite{Thornburg2003:AH-finding} to locate apparent
horizons and measure the magnitude of the horizon spin $S_H$, using
the {\it isolated horizon} algorithm as implemented in
Ref.~\cite{Campanelli:2006fy}.  We measure radiated energy, linear
momentum, and angular momentum, in terms of the radiative Weyl scalar
$\psi_4$, using the formulas provided in
Refs.~\cite{Campanelli:1998jv, Lousto:2007mh}.  As described in
Ref.~\cite{Nakano:2015pta}, we use the Teukolsky equation to
analytically extrapolate expressions for $R \psi_4$ from a finite
observer location ($R_{obs}>100M$) to infinity ($\scri^+$).

\begin{figure}
\includegraphics[angle=0,width=\columnwidth]{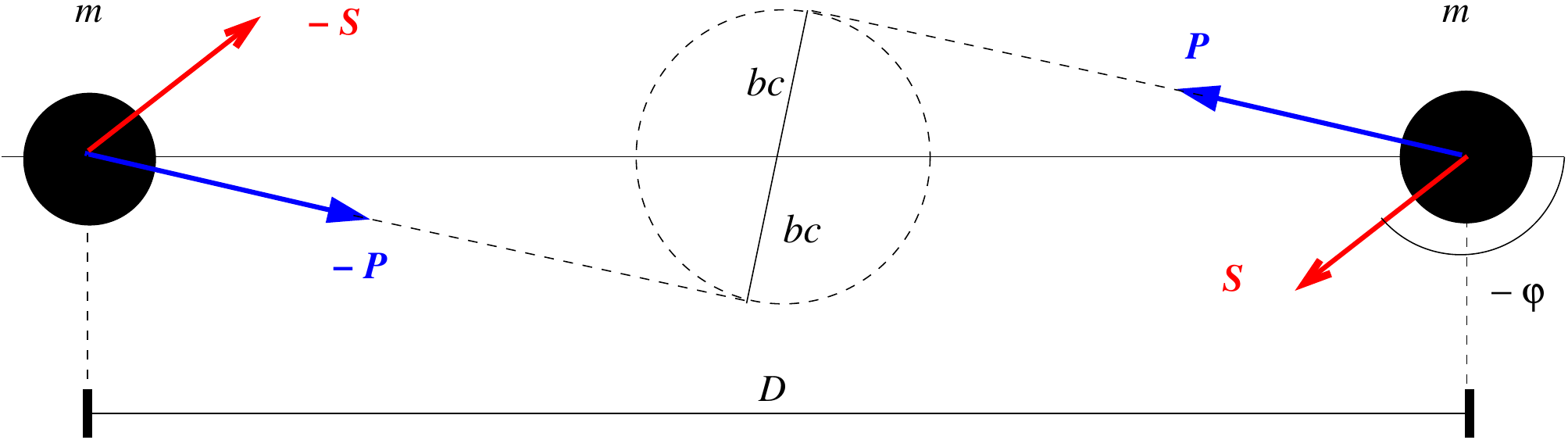}
\caption{Maximum high energy collision kicks binary black hole initial configurations.
  On the orbital plane equal mass $m$ black holes with opposing spin $\vec{S}$
  and momentum $\vec{P}$ with critical impact parameter $b_c$ and starting
  separation $D=50M$.
  \label{fig:IDConfig}}
\end{figure}

One can hypothesize on asymmetry properties that one can search for the
maximum recoil within a family of equal mass, opposite spins on the
orbital plane configurations, as displayed in Fig.~\ref{fig:IDConfig}.
The compromise with maximizing the energy radiated via the hangup
effect \cite{Campanelli:2006uy} that we needed to find the maximum
recoil for quasicircular orbits in \cite{Lousto:2011kp}
is replaced here by the determination of the critical impact
parameter, $b_c$, separating merger from scattering of the holes.
Besides, in Ref.~\cite{Sperhake:2012me} it is displayed a certain
independence of the energy radiated with the spin of the holes.

This Fig.~\ref{fig:IDConfig} configuration has been studied earlier in
\cite{Healy:2008js,Sperhake:2010uv} leading to a wide range of maximum
velocity estimates of 10,000 km/s and 15,000km/s from simulations to
potential extrapolations up to 45,000km/s.  Here we will study this problem
in detail with our new set of specially designed simulations to
explicitly model the problem in terms of the Bowen-York initial
momentum of the holes, $\gamma v$, impact parameter, $b$, and spin,
$\vec{s}=\vec{S}_H/m_H^2$ (where $m_H=m_{1,2}$ is the horizon mass of
each hole), i.e. a four dimensional parameter search.

\section{Simulations' Results\label{sec:results}}

\begin{figure}
\includegraphics[angle=0,width=0.85\columnwidth]{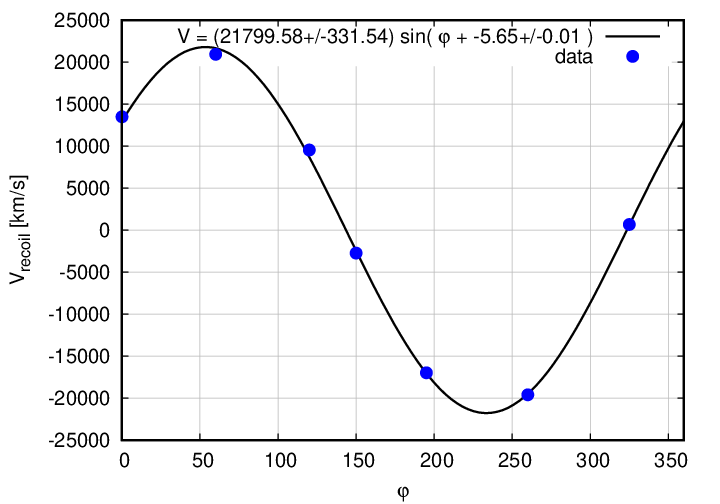}
  \caption{A series of simulations versus $\varphi$ orientation of the spin for  $b=2.38, s=0.85, \gamma v=0.874$.
  \label{fig:Vvsphi}}
\end{figure}

Our simulations families consist of a choice of an initial
(Bowen-York) data spin magnitude, here $s=0.4, 0.7, 0.8, 0.85, 0.9$,
and for each of them an initial momentum per irreducible mass, $\gamma
v$, and impact parameter, $bM$, as measured at the initial separation
of the holes $D=50M$ (with $M=m_1+m_2$ the addition of the horizon
masses of the system). We then vary the orientation of the spins
pointing on the orbital plane by an angle $\varphi$ with respect to
the line initially joining the black holes. This allows us to model
the leading $\varphi$-dependence of the recoil velocity as a
$\cos\varphi$ \cite{Lousto:2010xk}. In practice one needs about 4 to 7
simulations to fit this dependence and determine the amplitude of the
curve leading to the maximum recoil for this configuration, as for
instance displayed in Fig.~\ref{fig:Vvsphi}.

To compute recoil velocities from these waveforms in the time domain,
we subtract, at the post-processing stage, the initial burst of
spurious radiation. This Bowen-York initial data radiation content
reaches earlier the fiducial observer than its more physical
components, thus can be simply excised.
The effects of this spurious radiating is nevertheless relatively
much smaller on the recoil velocities than it is on
the radiated energy. For instance, for a case closest to maximum kick
$(s=0.9, b=2.38, \gamma v=1.1)$ we get a kick of about 22,700 km/s total,
with the spurious burst contributing around 25 km/s. While
for the energy radiated, the spurious burst contributes approximately
5\% $(0.018/0.336)M$.

\begin{figure}
\includegraphics[angle=0,width=\columnwidth]{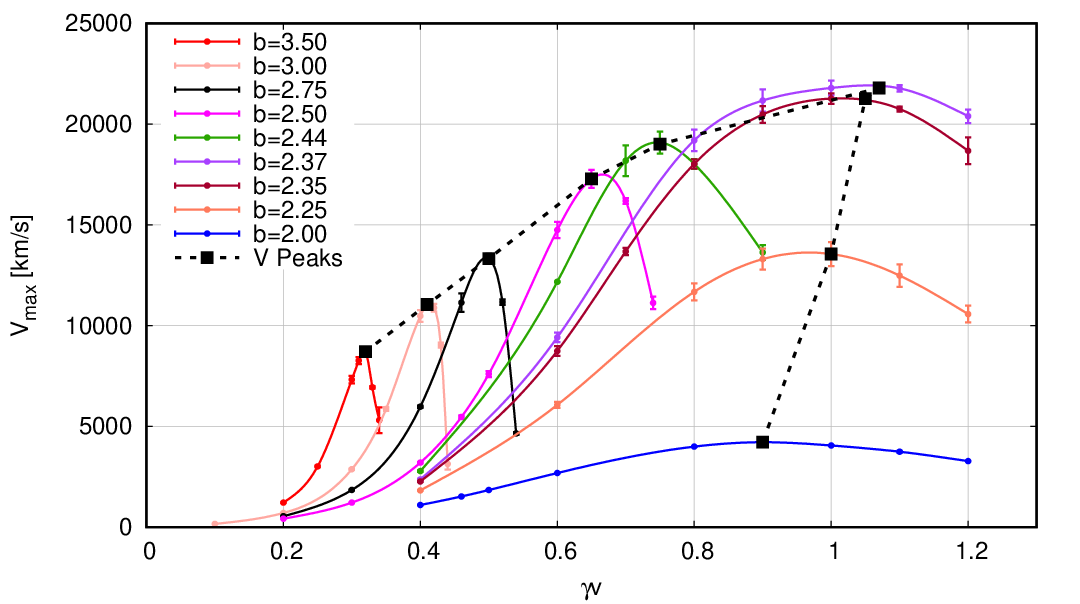}
\caption{A series of simulations versus $\gamma v$ and $b$ to search for
  $V_{\text{max}}$ for the spin $s=0.80$ case.
  \label{fig:Vvsgvb}}
\end{figure}

The next step in the search for the maximum recoil is to find the peak
velocities in the $\gamma v$ and $b$ parameter space. This is achieved
as displayed in Fig.~\ref{fig:Vvsgvb}.

\begin{figure}
\includegraphics[angle=0,width=0.9\columnwidth]{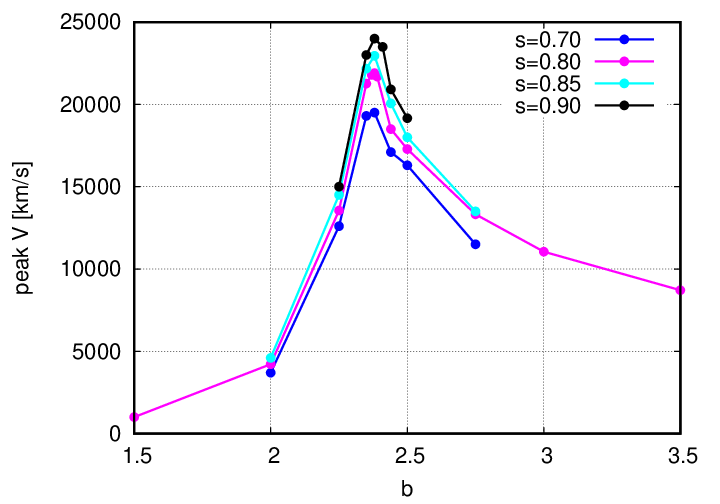}
  \caption{Maximum recoil velocity for different impact parameters $b$. For all the high spins studied here this peaks at $b\approx2.38$.
  \label{fig:Vvsb}}
\end{figure}

This search process of the impact parameter $b$ to find the value
$b_{\text{max}}$ leading to the largest recoil velocity is then
repeated for each spin value.  Fig.~\ref{fig:Vvsb} displays this
search for each of the spin magnitudes we considered here.  In
practice, the $b_{\text{max}}$ corresponds closely to the critical
value of the impact parameter $b_{c}$ separating the direct merger
from the scattering of the holes.

\begin{figure}
\includegraphics[angle=0,width=0.9\columnwidth]{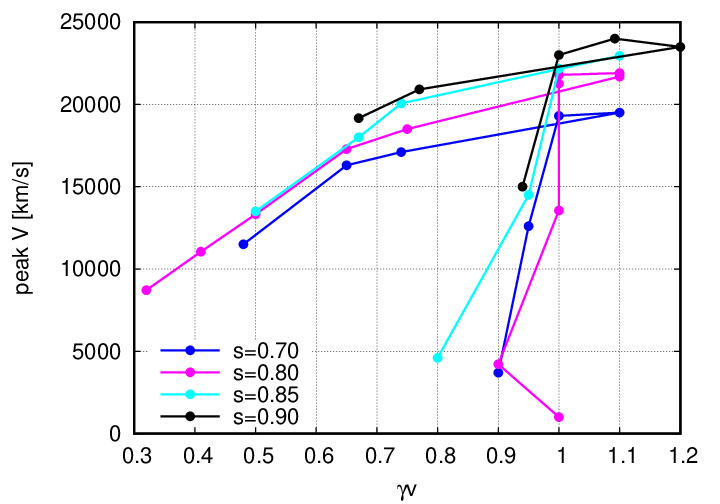}
  \caption{Display of peak velocity vs. $\gamma v$ and spin for merging holes. 
  \label{fig:Vvsgvvsa}}
\end{figure}

A similar analysis can be done by varying the initial velocity $v$, or
rather the linear momentum per irreducible mass of the holes, $\gamma
v=P/m_{irr}$, with $\gamma=(1-v^2)^{-1/2}$, the Lorentz factor, and
$A_H=16\pi m_{irr}^2$ the measured horizon area.  As displayed in
Fig.~\ref{fig:Vvsgvvsa} those curves display the same feature of
maximizing the recoil velocity for values about the critical momentum
separating the direct merger from scattering of the holes. The return
loop of the curves around its maximum represents an overshot of the
impact parameter compensated by the lowering of the initial velocity
to warrant merger instead of scattering.

\begin{figure}
\includegraphics[angle=0,width=\columnwidth]{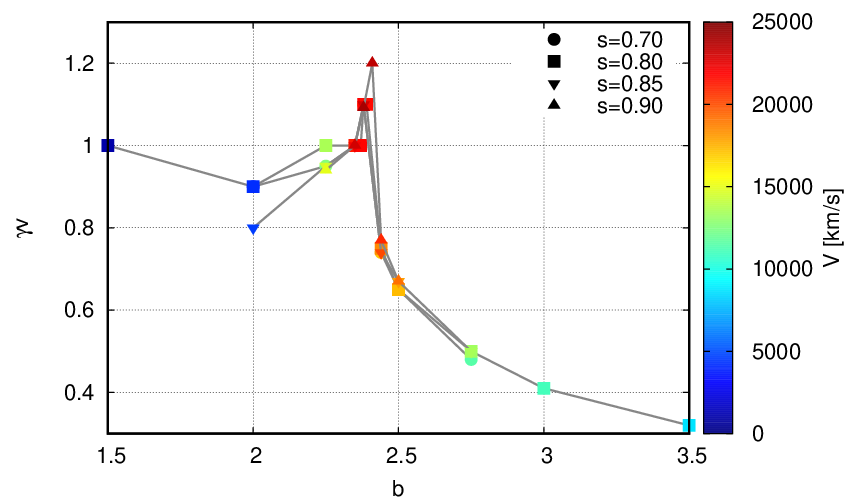}
  \caption{Maximum recoil velocity for different initial momenta parameters $\gamma v$ and impact parameters $b$ as a color map. 
  \label{fig:Vvsgv}}
\end{figure}

The explicit dependence on both parameters, $b$ and $\gamma v$ is displayed
in Fig.~\ref{fig:Vvsgv} as a heat map for each of the spin magnitudes
$s= 0.7, 0.8, 0.85, 0.9$.

\begin{figure}
\includegraphics[angle=0,width=\columnwidth]{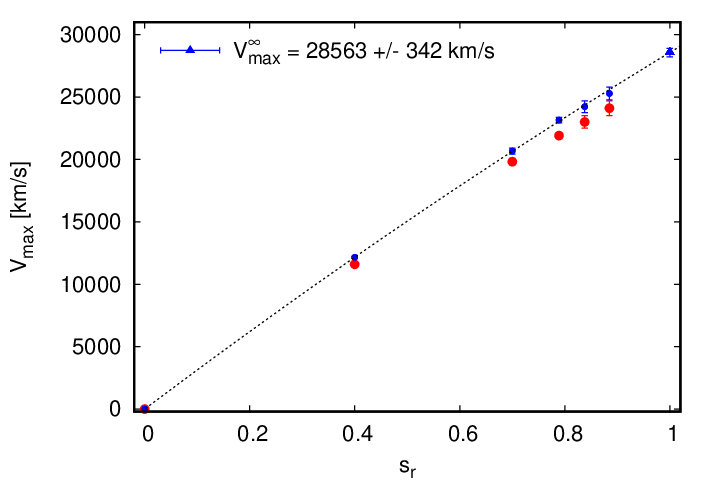}
\caption{Maximum recoil velocity versus the settled spins value $s_r$ and its extrapolation to maximal spin $s_r=1$.
(Blue points are extrapolation to infinite resolution, red points are the n100 low resolution results).
  \label{fig:Vmax}}
\end{figure}

The final results of the maximum recoil velocities for each $s$ and
the corresponding relaxed (at around $t=30M$) spin magnitude, $|s_r|$,
are summarized in Table~\ref{tab:ID+} and in Fig.~\ref{fig:Vmax},
where we display the error bars of each point and fit to a quadratic
dependence fit on $s_r$ to extrapolate to the ultimate recoil velocity
finding $28,562\pm342$ km/s for the extremely spinning, $s_r=1$,
binary black holes case. We also display for comparison with the
extrapolated to infinite resolution values (in blue) used for the
fit, the n100 low resolution results (in red), used for the parameter
searches.

\begin{table*}
  \caption{All simulations have equal mass $m_1=m_2$ black holes, and
    are initially placed at $x_{1,2}=\pm25M$. The relaxed spin
    magnitudes, $|s_r|$, are used for the final fit.  Measured maximal
    recoil velocities and its extrapolation (order) to infinite
    resolution are given on the right panel.  }\label{tab:ID+}
\begin{ruledtabular}
\begin{tabular}{rcccc|ccccc}
$\#$ runs & $\pm s$  & $|s_r|$ & $b_{\text{max}}$ & $(\gamma v)_{\text{max}}$  & $V_{\text{max}}^{n100}$ [km/s] & $V_{\text{max}}^{n120}$ [km/s] & $V_{\text{max}}^{n144}$ [km/s] & $V_{\text{max}}^\infty$ [km/s] & order \\
  \hline
72 &  0.40 & 0.400 & 2.38 & 1.20 & 11,637 $\pm$  67 & 11,827 $\pm$ 67  & 11,944 $\pm$ 64  & 12,133 $\pm$ 189 & 2.7\\
233&  0.70 & 0.699 & 2.38 & 1.10 & 19,832 $\pm$ 267 & 20,163 $\pm$ 267 & 20,360 $\pm$ 262 & 20,649 $\pm$ 289 & 2.9\\
472&  0.80 & 0.789 & 2.38 & 1.10 & 22,212 $\pm$ 228 & 22,583 $\pm$ 226 & 22,800 $\pm$ 217 & 23,104 $\pm$ 304 & 3.0\\
305&  0.85 & 0.838 & 2.38 & 1.10 & 23,291 $\pm$ 514 & 23,666 $\pm$ 486 & 23,892 $\pm$ 482 & 24,231 $\pm$ 339 & 2.8\\ 
299&  0.90 & 0.885 & 2.38 & 1.09 & 24,172 $\pm$ 579 & 24,609 $\pm$ 565 & 24,870 $\pm$ 552 & 25,256 $\pm$ 386 & 2.8\\
\end{tabular}
\end{ruledtabular}
\end{table*}

Note also that in Table~\ref{tab:ID+} we provide the number of runs
(simulations) we performed in the three-dimensional search
($\varphi,b,\gamma v$) of the maximum recoil for a given spin $s$. In
the case of $s=0.4$, to simplify the search, and on the light of the
previous results with the higher spin cases (as seen in
Figs.~\ref{fig:Vvsb} and \ref{fig:Vvsgv}) we assumed $b=2.38$,
hence the lower number of simulations needed.

For the majority of the simulations, for spins of $s=0.4$ to $s=0.85$,
we use a grid, labeled as n100, with 10 levels of refinement, the
coarsest of which has resolution of $4M$ and outer boundary of
$r=400M$, with each successive grid with twice the resolution. If we
label the coarsest grid $n=0$, and the finest grid $n=9$, the
resolution on a given level is $M/2^{(n-2)}$.  The wavezone is $n=2$
with a resolution of $M/1$ and boundary out to $r=125M$.  The finest
grid has a resolution of $M/128$ with a size of $0.5M$ centered around
each black hole.  The spin $s=0.9$ case has an additional refinement
level around each black hole with resolution $M/256$ and a radius of
$r=0.3M$.

To evaluate the finite differences errors and extrapolation of our
simulations, we have performed three simulation sets with increasing
global resolutions by factors of 1.2 (n120, n144) with respect to our
base resolution, n100, for the peak velocity cases with $b=2.38$,
$(\gamma v)_{\text{max}}$, and four
$\varphi=0^\circ,60^\circ,120^\circ,150^\circ$ degrees for each of the
spins $s=0.40, 0.70, 0.80, 0.85, 0.90$. The resulting measured recoil
velocities are given in Table~\ref{tab:ID+}. Extrapolation to
Infinite resolution leads to $V^{\infty}_{\text{max}}$ values representing
about a $3\%$ increase from the n100 results.  The 3rd order
convergence rate found for the net recoil (computed as large
differences of anisotropic radiation), is what one expects from the
4th Runge-Kutta time integrator used by our code.

As a further check of our numerical accuracy, we have recalculated a
set of cases for the spin 0.8 and 0.85 with the extra refinement level
and increased grid sizes as we used for the spin 0.9 runs. We then
recalculated the series that gives a maximum value for spin 0.8 of
$21,802\pm191$ km/s. Compared to the original grid computation of
$21,903\pm213$ km/s, this leads to a difference of 101 km/s or
$0.46\%$.

\section{Conclusions\label{sec:conclusions}}

\begin{figure}
\includegraphics[angle=0,width=\columnwidth]{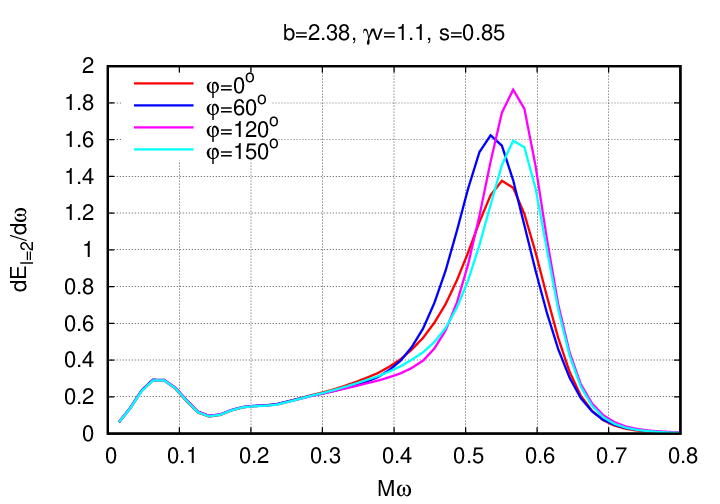}
\caption{The spectrum of the ($\ell=2$-modes) energy radiated
  $dE_{\ell=2}/d\omega$ by a representative set of simulations (with
  $b=2.38$, $s=0.85$, $\gamma v=1.1$) for different orientation
  angles, $\varphi$, of the spin.
  \label{fig:dE2_dw}}
\end{figure}

In Fig.~\ref{fig:dE2_dw} we display the spectrum of radiated energy by
adding the leading ($\ell=2, m=0,\pm2$)-modes for one of the peak
recoil cases ($b=2.38$, $s=0.85$, $\gamma v=1.1$) for different
orientation angles, $\varphi$, of the spin. We observe a bulge at low
frequencies, corresponding to the initial data content and
``bremsstrahlung-like'' radiation of the holes approaching each other
from $D=50M$ and that the different spin orientations do not produce
notable differences in this part of the spectrum.  Meanwhile, at
higher frequencies (by an order of magnitude), corresponding to when
the holes reach the critical separation $2b_cM=4.76M$ and then
subsequently merge, the spectrum shows a strong dependence on the spin
orientations.

To summarize, we have been able to provide an accurate estimate of the
ultimate recoil, product of the high energy collision of two black
holes. In order to perform the four dimensional search (momentum
$\gamma v$, impact parameter $b$, spin orientation $\varphi$ and
magnitude $s$) we performed 1381 simulations in search for the
critical $b_c$ marginally leading to merger and the corresponding
value of $\gamma_{\text{max}}$ that maximized the recoil all as a
function of $\varphi$ for each $s$.  Extrapolation to extreme spins
(as shown in Fig.~\ref{fig:Vmax}) have led us to estimate the value of
$28,562\pm 342$ km/s for the ultimate recoil, placing thus a bound
for it of below 10\% the speed of light.

We thus note here the crucial relevance of the holes' spin magnitude
and orientation in the determination of the high energy collision
kicks. In a follow up paper~\cite{CHLinprep} we plan to study in
detail also the role of the spins in the determination of the absolute
maximum energy and angular momentum radiated by such systems.


Figs. \ref{fig:Vvsb} and \ref{fig:Vvsgv} display a cusp like
dependence with the impact parameter $b$ around its $b_{\text{max}}$
value that is reminiscent of critical behavior separating direct
merger from scattering of the holes.  Critical behavior was discussed
in detail in Ref.~\cite{Berti:2010ce} in the context of the energy
radiated by this ultrarelativistic black hole encounters, identifying
impacts parameters for direct merger $b^*$ and unbound scattering
$b_{\text{scat}}$.  Here we observe this kind of behavior regarding
the approach to the maximum radiated {\it linear} momentum as the
impact parameter reaches its critical value $b_{\text{max}}$ and
$\gamma v_{\text{max}}$, analogous to the critical temperature and
pressure in a liquid-gas system. The intermediate impact parameter
region between direct merger and direct scattering acts in a way to
match those values of the recoil velocity in a continuous way,
thus conforming the analogous of a second order phase transition (in
the Ehrenfest classification).  Further studies to verify critical
behavior and to obtain critical exponents, as to verify to what extent
one can speak of universality, order parameters and scaling properties
in this context, is certainly of interest and deserves follow up
research, preferably with semianalytic methods \cite{CHLinprep} since
3D critical phenomena remains extremely challenging with full
numerical relativity.

\begin{acknowledgments}
The authors gratefully acknowledge the National Science Foundation
(NSF) for financial support from Grant No.\ PHY-1912632 and PHY-2207920.
Computational resources were also provided by the New Horizons, Blue
Sky, Green Prairies, and White Lagoon clusters at the CCRG-Rochester
Institute of Technology, which were supported by NSF grants
No.\ PHY-0722703, No.\ DMS-0820923, No.\ AST-1028087,
No.\ PHY-1229173, No.\ PHY-1726215, and No.\ PHY-2018420.  This work
used the 
ACCESS allocation TG-PHY060027N, founded by NSF,
and project PHY20007 Frontera, an NSF-funded
Petascale computing system at the Texas Advanced Computing Center
(TACC).
\end{acknowledgments}

\bibliography{../../../../../Bibtex/references}

\begin{thebibliography}{48}%
\makeatletter
\providecommand \@ifxundefined [1]{%
 \@ifx{#1\undefined}
}%
\providecommand \@ifnum [1]{%
 \ifnum #1\expandafter \@firstoftwo
 \else \expandafter \@secondoftwo
 \fi
}%
\providecommand \@ifx [1]{%
 \ifx #1\expandafter \@firstoftwo
 \else \expandafter \@secondoftwo
 \fi
}%
\providecommand \natexlab [1]{#1}%
\providecommand \enquote  [1]{``#1''}%
\providecommand \bibnamefont  [1]{#1}%
\providecommand \bibfnamefont [1]{#1}%
\providecommand \citenamefont [1]{#1}%
\providecommand \href@noop [0]{\@secondoftwo}%
\providecommand \href [0]{\begingroup \@sanitize@url \@href}%
\providecommand \@href[1]{\@@startlink{#1}\@@href}%
\providecommand \@@href[1]{\endgroup#1\@@endlink}%
\providecommand \@sanitize@url [0]{\catcode `\\12\catcode `\$12\catcode
  `\&12\catcode `\#12\catcode `\^12\catcode `\_12\catcode `\%12\relax}%
\providecommand \@@startlink[1]{}%
\providecommand \@@endlink[0]{}%
\providecommand \url  [0]{\begingroup\@sanitize@url \@url }%
\providecommand \@url [1]{\endgroup\@href {#1}{\urlprefix }}%
\providecommand \urlprefix  [0]{URL }%
\providecommand \Eprint [0]{\href }%
\providecommand \doibase [0]{http://dx.doi.org/}%
\providecommand \selectlanguage [0]{\@gobble}%
\providecommand \bibinfo  [0]{\@secondoftwo}%
\providecommand \bibfield  [0]{\@secondoftwo}%
\providecommand \translation [1]{[#1]}%
\providecommand \BibitemOpen [0]{}%
\providecommand \bibitemStop [0]{}%
\providecommand \bibitemNoStop [0]{.\EOS\space}%
\providecommand \EOS [0]{\spacefactor3000\relax}%
\providecommand \BibitemShut  [1]{\csname bibitem#1\endcsname}%
\let\auto@bib@innerbib\@empty
\bibitem [{\citenamefont {Campanelli}\ \emph
  {et~al.}(2007{\natexlab{a}})\citenamefont {Campanelli}, \citenamefont
  {Lousto}, \citenamefont {Zlochower},\ and\ \citenamefont
  {Merritt}}]{Campanelli:2007ew}%
  \BibitemOpen
  \bibfield  {author} {\bibinfo {author} {\bibfnamefont {M.}~\bibnamefont
  {Campanelli}}, \bibinfo {author} {\bibfnamefont {C.~O.}\ \bibnamefont
  {Lousto}}, \bibinfo {author} {\bibfnamefont {Y.}~\bibnamefont {Zlochower}}, \
  and\ \bibinfo {author} {\bibfnamefont {D.}~\bibnamefont {Merritt}},\
  }\href@noop {} {\bibfield  {journal} {\bibinfo  {journal} {Astrophys. J.}\
  }\textbf {\bibinfo {volume} {659}},\ \bibinfo {pages} {L5} (\bibinfo {year}
  {2007}{\natexlab{a}})},\ \Eprint {http://arxiv.org/abs/gr-qc/0701164}
  {gr-qc/0701164} \BibitemShut {NoStop}%
\bibitem [{\citenamefont {Gonz\'alez}\ \emph {et~al.}(2007)\citenamefont
  {Gonz\'alez}, \citenamefont {Hannam}, \citenamefont {Sperhake}, \citenamefont
  {Brugmann},\ and\ \citenamefont {Husa}}]{Gonzalez:2007hi}%
  \BibitemOpen
  \bibfield  {author} {\bibinfo {author} {\bibfnamefont {J.~A.}\ \bibnamefont
  {Gonz\'alez}}, \bibinfo {author} {\bibfnamefont {M.~D.}\ \bibnamefont
  {Hannam}}, \bibinfo {author} {\bibfnamefont {U.}~\bibnamefont {Sperhake}},
  \bibinfo {author} {\bibfnamefont {B.}~\bibnamefont {Brugmann}}, \ and\
  \bibinfo {author} {\bibfnamefont {S.}~\bibnamefont {Husa}},\ }\href@noop {}
  {\bibfield  {journal} {\bibinfo  {journal} {Phys. Rev. Lett.}\ }\textbf
  {\bibinfo {volume} {98}},\ \bibinfo {pages} {231101} (\bibinfo {year}
  {2007})},\ \Eprint {http://arxiv.org/abs/gr-qc/0702052} {gr-qc/0702052}
  \BibitemShut {NoStop}%
\bibitem [{\citenamefont {Komossa}(2012)}]{Komossa:2012cy}%
  \BibitemOpen
  \bibfield  {author} {\bibinfo {author} {\bibfnamefont {S.}~\bibnamefont
  {Komossa}},\ }\href@noop {} {\bibfield  {journal} {\bibinfo  {journal} {Adv.
  Astron.}\ }\textbf {\bibinfo {volume} {2012}},\ \bibinfo {pages} {364973}
  (\bibinfo {year} {2012})},\ \Eprint {http://arxiv.org/abs/1202.1977}
  {arXiv:1202.1977 [astro-ph.CO]} \BibitemShut {NoStop}%
\bibitem [{\citenamefont {Chiaberge}\ \emph {et~al.}(2018)\citenamefont
  {Chiaberge}, \citenamefont {Tremblay}, \citenamefont {Capetti},\ and\
  \citenamefont {Norman}}]{Chiaberge:2018lkg}%
  \BibitemOpen
  \bibfield  {author} {\bibinfo {author} {\bibfnamefont {M.}~\bibnamefont
  {Chiaberge}}, \bibinfo {author} {\bibfnamefont {G.~R.}\ \bibnamefont
  {Tremblay}}, \bibinfo {author} {\bibfnamefont {A.}~\bibnamefont {Capetti}}, \
  and\ \bibinfo {author} {\bibfnamefont {C.}~\bibnamefont {Norman}},\ }\href
  {\doibase 10.3847/1538-4357/aac48b} {\bibfield  {journal} {\bibinfo
  {journal} {Astrophys. J.}\ }\textbf {\bibinfo {volume} {861}},\ \bibinfo
  {pages} {56} (\bibinfo {year} {2018})},\ \Eprint
  {http://arxiv.org/abs/1805.05860} {arXiv:1805.05860 [astro-ph.GA]}
  \BibitemShut {NoStop}%
\bibitem [{\citenamefont {Campanelli}\ \emph
  {et~al.}(2007{\natexlab{b}})\citenamefont {Campanelli}, \citenamefont
  {Lousto}, \citenamefont {Zlochower},\ and\ \citenamefont
  {Merritt}}]{Campanelli:2007cga}%
  \BibitemOpen
  \bibfield  {author} {\bibinfo {author} {\bibfnamefont {M.}~\bibnamefont
  {Campanelli}}, \bibinfo {author} {\bibfnamefont {C.~O.}\ \bibnamefont
  {Lousto}}, \bibinfo {author} {\bibfnamefont {Y.}~\bibnamefont {Zlochower}}, \
  and\ \bibinfo {author} {\bibfnamefont {D.}~\bibnamefont {Merritt}},\
  }\href@noop {} {\bibfield  {journal} {\bibinfo  {journal} {Phys. Rev. Lett.}\
  }\textbf {\bibinfo {volume} {98}},\ \bibinfo {pages} {231102} (\bibinfo
  {year} {2007}{\natexlab{b}})},\ \Eprint {http://arxiv.org/abs/gr-qc/0702133}
  {gr-qc/0702133} \BibitemShut {NoStop}%
\bibitem [{\citenamefont {Lousto}\ and\ \citenamefont
  {Zlochower}(2011{\natexlab{a}})}]{Lousto:2011kp}%
  \BibitemOpen
  \bibfield  {author} {\bibinfo {author} {\bibfnamefont {C.~O.}\ \bibnamefont
  {Lousto}}\ and\ \bibinfo {author} {\bibfnamefont {Y.}~\bibnamefont
  {Zlochower}},\ }\href {\doibase 10.1103/PhysRevLett.107.231102} {\bibfield
  {journal} {\bibinfo  {journal} {Phys. Rev. Lett.}\ }\textbf {\bibinfo
  {volume} {107}},\ \bibinfo {pages} {231102} (\bibinfo {year}
  {2011}{\natexlab{a}})},\ \Eprint {http://arxiv.org/abs/1108.2009}
  {arXiv:1108.2009 [gr-qc]} \BibitemShut {NoStop}%
\bibitem [{\citenamefont {Campanelli}\ \emph
  {et~al.}(2006{\natexlab{a}})\citenamefont {Campanelli}, \citenamefont
  {Lousto},\ and\ \citenamefont {Zlochower}}]{Campanelli:2006uy}%
  \BibitemOpen
  \bibfield  {author} {\bibinfo {author} {\bibfnamefont {M.}~\bibnamefont
  {Campanelli}}, \bibinfo {author} {\bibfnamefont {C.~O.}\ \bibnamefont
  {Lousto}}, \ and\ \bibinfo {author} {\bibfnamefont {Y.}~\bibnamefont
  {Zlochower}},\ }\href@noop {} {\bibfield  {journal} {\bibinfo  {journal}
  {Phys. Rev.}\ }\textbf {\bibinfo {volume} {D74}},\ \bibinfo {pages}
  {041501(R)} (\bibinfo {year} {2006}{\natexlab{a}})},\ \Eprint
  {http://arxiv.org/abs/gr-qc/0604012} {gr-qc/0604012} \BibitemShut {NoStop}%
\bibitem [{\citenamefont {Gayathri}\ \emph {et~al.}(2022)\citenamefont
  {Gayathri}, \citenamefont {Healy}, \citenamefont {Lange}, \citenamefont
  {O'Brien}, \citenamefont {Szczepanczyk}, \citenamefont {Bartos},
  \citenamefont {Campanelli}, \citenamefont {Klimenko}, \citenamefont
  {Lousto},\ and\ \citenamefont {O'Shaughnessy}}]{Gayathri:2020coq}%
  \BibitemOpen
  \bibfield  {author} {\bibinfo {author} {\bibfnamefont {V.}~\bibnamefont
  {Gayathri}}, \bibinfo {author} {\bibfnamefont {J.}~\bibnamefont {Healy}},
  \bibinfo {author} {\bibfnamefont {J.}~\bibnamefont {Lange}}, \bibinfo
  {author} {\bibfnamefont {B.}~\bibnamefont {O'Brien}}, \bibinfo {author}
  {\bibfnamefont {M.}~\bibnamefont {Szczepanczyk}}, \bibinfo {author}
  {\bibfnamefont {I.}~\bibnamefont {Bartos}}, \bibinfo {author} {\bibfnamefont
  {M.}~\bibnamefont {Campanelli}}, \bibinfo {author} {\bibfnamefont
  {S.}~\bibnamefont {Klimenko}}, \bibinfo {author} {\bibfnamefont {C.~O.}\
  \bibnamefont {Lousto}}, \ and\ \bibinfo {author} {\bibfnamefont
  {R.}~\bibnamefont {O'Shaughnessy}},\ }\href {\doibase
  10.1038/s41550-021-01568-w} {\bibfield  {journal} {\bibinfo  {journal}
  {Nature Astron.}\ }\textbf {\bibinfo {volume} {6}},\ \bibinfo {pages} {344}
  (\bibinfo {year} {2022})},\ \Eprint {http://arxiv.org/abs/2009.05461}
  {arXiv:2009.05461 [astro-ph.HE]} \BibitemShut {NoStop}%
\bibitem [{\citenamefont {Cardoso}\ \emph {et~al.}(2012)\citenamefont
  {Cardoso}, \citenamefont {Gualtieri}, \citenamefont {Herdeiro}, \citenamefont
  {Sperhak}, \citenamefont {Chesler}, \citenamefont {Lehner}, \citenamefont
  {Park}, \citenamefont {Reall}, \citenamefont {(section coordinators)},
  \citenamefont {Alic}, \citenamefont {Dias}, \citenamefont {Emparan},
  \citenamefont {Ferrari}, \citenamefont {Giddings}, \citenamefont {Godazgar},
  \citenamefont {Gregory}, \citenamefont {Hubeny}, \citenamefont {Ishibashi},
  \citenamefont {Landsberg}, \citenamefont {Lousto}, \citenamefont {Mateos},
  \citenamefont {Moeller}, \citenamefont {Okawa}, \citenamefont {Pani},
  \citenamefont {Parker}, \citenamefont {Pretorius}, \citenamefont {Shibata},
  \citenamefont {Sotani}, \citenamefont {Wiseman}, \citenamefont {Witek},
  \citenamefont {Yunes},\ and\ \citenamefont {Zilh{\~a}o}}]{Cardoso:2012qm}%
  \BibitemOpen
  \bibfield  {author} {\bibinfo {author} {\bibfnamefont {V.}~\bibnamefont
  {Cardoso}}, \bibinfo {author} {\bibfnamefont {L.}~\bibnamefont {Gualtieri}},
  \bibinfo {author} {\bibfnamefont {C.}~\bibnamefont {Herdeiro}}, \bibinfo
  {author} {\bibfnamefont {U.}~\bibnamefont {Sperhak}}, \bibinfo {author}
  {\bibfnamefont {P.~M.}\ \bibnamefont {Chesler}}, \bibinfo {author}
  {\bibfnamefont {L.}~\bibnamefont {Lehner}}, \bibinfo {author} {\bibfnamefont
  {S.~C.}\ \bibnamefont {Park}}, \bibinfo {author} {\bibfnamefont {H.~S.}\
  \bibnamefont {Reall}}, \bibinfo {author} {\bibfnamefont {C.~F.~S.}\
  \bibnamefont {(section coordinators)}}, \bibinfo {author} {\bibfnamefont
  {D.}~\bibnamefont {Alic}}, \bibinfo {author} {\bibfnamefont {O.~J.~C.}\
  \bibnamefont {Dias}}, \bibinfo {author} {\bibfnamefont {R.}~\bibnamefont
  {Emparan}}, \bibinfo {author} {\bibfnamefont {V.}~\bibnamefont {Ferrari}},
  \bibinfo {author} {\bibfnamefont {S.~B.}\ \bibnamefont {Giddings}}, \bibinfo
  {author} {\bibfnamefont {M.}~\bibnamefont {Godazgar}}, \bibinfo {author}
  {\bibfnamefont {R.}~\bibnamefont {Gregory}}, \bibinfo {author} {\bibfnamefont
  {V.~E.}\ \bibnamefont {Hubeny}}, \bibinfo {author} {\bibfnamefont
  {A.}~\bibnamefont {Ishibashi}}, \bibinfo {author} {\bibfnamefont
  {G.}~\bibnamefont {Landsberg}}, \bibinfo {author} {\bibfnamefont {C.~O.}\
  \bibnamefont {Lousto}}, \bibinfo {author} {\bibfnamefont {D.}~\bibnamefont
  {Mateos}}, \bibinfo {author} {\bibfnamefont {V.}~\bibnamefont {Moeller}},
  \bibinfo {author} {\bibfnamefont {H.}~\bibnamefont {Okawa}}, \bibinfo
  {author} {\bibfnamefont {P.}~\bibnamefont {Pani}}, \bibinfo {author}
  {\bibfnamefont {M.~A.}\ \bibnamefont {Parker}}, \bibinfo {author}
  {\bibfnamefont {F.}~\bibnamefont {Pretorius}}, \bibinfo {author}
  {\bibfnamefont {M.}~\bibnamefont {Shibata}}, \bibinfo {author} {\bibfnamefont
  {H.}~\bibnamefont {Sotani}}, \bibinfo {author} {\bibfnamefont
  {T.}~\bibnamefont {Wiseman}}, \bibinfo {author} {\bibfnamefont
  {H.}~\bibnamefont {Witek}}, \bibinfo {author} {\bibfnamefont
  {N.}~\bibnamefont {Yunes}}, \ and\ \bibinfo {author} {\bibfnamefont
  {M.}~\bibnamefont {Zilh{\~a}o}},\ }\href {\doibase
  10.1088/0264-9381/29/24/244001} {\bibfield  {journal} {\bibinfo  {journal}
  {Class. Quant. Grav.}\ }\textbf {\bibinfo {volume} {29}},\ \bibinfo {pages}
  {244001} (\bibinfo {year} {2012})},\ \Eprint {http://arxiv.org/abs/1201.5118}
  {arXiv:1201.5118 [hep-th]} \BibitemShut {NoStop}%
\bibitem [{\citenamefont {Berti}\ \emph {et~al.}(2016)\citenamefont {Berti},
  \citenamefont {Cardoso}, \citenamefont {Crispino}, \citenamefont {Gualtieri},
  \citenamefont {Herdeiro},\ and\ \citenamefont {Sperhake}}]{Berti:2016rij}%
  \BibitemOpen
  \bibfield  {author} {\bibinfo {author} {\bibfnamefont {E.}~\bibnamefont
  {Berti}}, \bibinfo {author} {\bibfnamefont {V.}~\bibnamefont {Cardoso}},
  \bibinfo {author} {\bibfnamefont {L.~C.~B.}\ \bibnamefont {Crispino}},
  \bibinfo {author} {\bibfnamefont {L.}~\bibnamefont {Gualtieri}}, \bibinfo
  {author} {\bibfnamefont {C.}~\bibnamefont {Herdeiro}}, \ and\ \bibinfo
  {author} {\bibfnamefont {U.}~\bibnamefont {Sperhake}},\ }\href {\doibase
  10.1142/S0218271816410224} {\bibfield  {journal} {\bibinfo  {journal} {Int.
  J. Mod. Phys. D}\ }\textbf {\bibinfo {volume} {25}},\ \bibinfo {pages}
  {1641022} (\bibinfo {year} {2016})},\ \Eprint
  {http://arxiv.org/abs/1603.06146} {arXiv:1603.06146 [gr-qc]} \BibitemShut
  {NoStop}%
\bibitem [{\citenamefont {Cardoso}\ \emph {et~al.}(2015)\citenamefont
  {Cardoso}, \citenamefont {Gualtieri}, \citenamefont {Herdeiro},\ and\
  \citenamefont {Sperhake}}]{Cardoso:2014uka}%
  \BibitemOpen
  \bibfield  {author} {\bibinfo {author} {\bibfnamefont {V.}~\bibnamefont
  {Cardoso}}, \bibinfo {author} {\bibfnamefont {L.}~\bibnamefont {Gualtieri}},
  \bibinfo {author} {\bibfnamefont {C.}~\bibnamefont {Herdeiro}}, \ and\
  \bibinfo {author} {\bibfnamefont {U.}~\bibnamefont {Sperhake}},\ }\href
  {\doibase 10.1007/lrr-2015-1} {\bibfield  {journal} {\bibinfo  {journal}
  {Living Rev. Relativity}\ }\textbf {\bibinfo {volume} {18}},\ \bibinfo
  {pages} {1} (\bibinfo {year} {2015})},\ \Eprint
  {http://arxiv.org/abs/1409.0014} {arXiv:1409.0014 [gr-qc]} \BibitemShut
  {NoStop}%
\bibitem [{\citenamefont {Franciolini}\ \emph {et~al.}(2022)\citenamefont
  {Franciolini}, \citenamefont {Maharana},\ and\ \citenamefont
  {Muia}}]{Franciolini:2022htd}%
  \BibitemOpen
  \bibfield  {author} {\bibinfo {author} {\bibfnamefont {G.}~\bibnamefont
  {Franciolini}}, \bibinfo {author} {\bibfnamefont {A.}~\bibnamefont
  {Maharana}}, \ and\ \bibinfo {author} {\bibfnamefont {F.}~\bibnamefont
  {Muia}},\ }\href {\doibase 10.1103/PhysRevD.106.103520} {\bibfield  {journal}
  {\bibinfo  {journal} {Phys. Rev. D}\ }\textbf {\bibinfo {volume} {106}},\
  \bibinfo {pages} {103520} (\bibinfo {year} {2022})},\ \Eprint
  {http://arxiv.org/abs/2205.02153} {arXiv:2205.02153 [astro-ph.CO]}
  \BibitemShut {NoStop}%
\bibitem [{\citenamefont {Ding}(2021)}]{Ding:2020ykt}%
  \BibitemOpen
  \bibfield  {author} {\bibinfo {author} {\bibfnamefont {Q.}~\bibnamefont
  {Ding}},\ }\href {\doibase 10.1103/PhysRevD.104.043527} {\bibfield  {journal}
  {\bibinfo  {journal} {Phys. Rev. D}\ }\textbf {\bibinfo {volume} {104}},\
  \bibinfo {pages} {043527} (\bibinfo {year} {2021})},\ \Eprint
  {http://arxiv.org/abs/2011.13643} {arXiv:2011.13643 [astro-ph.CO]}
  \BibitemShut {NoStop}%
\bibitem [{\citenamefont {Cai}\ and\ \citenamefont {Wang}(2020)}]{Cai:2019igo}%
  \BibitemOpen
  \bibfield  {author} {\bibinfo {author} {\bibfnamefont {R.-G.}\ \bibnamefont
  {Cai}}\ and\ \bibinfo {author} {\bibfnamefont {S.-J.}\ \bibnamefont {Wang}},\
  }\href {\doibase 10.1103/PhysRevD.101.043508} {\bibfield  {journal} {\bibinfo
   {journal} {Phys. Rev. D}\ }\textbf {\bibinfo {volume} {101}},\ \bibinfo
  {pages} {043508} (\bibinfo {year} {2020})},\ \Eprint
  {http://arxiv.org/abs/1910.07981} {arXiv:1910.07981 [astro-ph.CO]}
  \BibitemShut {NoStop}%
\bibitem [{\citenamefont {Hawking}(1971)}]{Hawking:1971tu}%
  \BibitemOpen
  \bibfield  {author} {\bibinfo {author} {\bibfnamefont {S.~W.}\ \bibnamefont
  {Hawking}},\ }\href {\doibase 10.1103/PhysRevLett.26.1344} {\bibfield
  {journal} {\bibinfo  {journal} {Phys. Rev. Lett.}\ }\textbf {\bibinfo
  {volume} {26}},\ \bibinfo {pages} {1344} (\bibinfo {year}
  {1971})}\BibitemShut {NoStop}%
\bibitem [{\citenamefont {Eardley}\ and\ \citenamefont
  {Giddings}(2002)}]{Eardley:2002re}%
  \BibitemOpen
  \bibfield  {author} {\bibinfo {author} {\bibfnamefont {D.~M.}\ \bibnamefont
  {Eardley}}\ and\ \bibinfo {author} {\bibfnamefont {S.~B.}\ \bibnamefont
  {Giddings}},\ }\href {\doibase 10.1103/PhysRevD.66.044011} {\bibfield
  {journal} {\bibinfo  {journal} {Phys. Rev.}\ }\textbf {\bibinfo {volume}
  {D66}},\ \bibinfo {pages} {044011} (\bibinfo {year} {2002})},\ \Eprint
  {http://arxiv.org/abs/gr-qc/0201034} {arXiv:gr-qc/0201034 [gr-qc]}
  \BibitemShut {NoStop}%
\bibitem [{\citenamefont {Siino}(2013)}]{Siino:2009vw}%
  \BibitemOpen
  \bibfield  {author} {\bibinfo {author} {\bibfnamefont {M.}~\bibnamefont
  {Siino}},\ }\href {\doibase 10.1142/S0218271813500508} {\bibfield  {journal}
  {\bibinfo  {journal} {Int.J.Mod.Phys.}\ }\textbf {\bibinfo {volume} {D22}},\
  \bibinfo {pages} {1350050} (\bibinfo {year} {2013})},\ \Eprint
  {http://arxiv.org/abs/0909.4827} {arXiv:0909.4827 [gr-qc]} \BibitemShut
  {NoStop}%
\bibitem [{\citenamefont {Mack}\ \emph {et~al.}(2007)\citenamefont {Mack},
  \citenamefont {Ostriker},\ and\ \citenamefont {Ricotti}}]{Mack:2006gz}%
  \BibitemOpen
  \bibfield  {author} {\bibinfo {author} {\bibfnamefont {K.~J.}\ \bibnamefont
  {Mack}}, \bibinfo {author} {\bibfnamefont {J.~P.}\ \bibnamefont {Ostriker}},
  \ and\ \bibinfo {author} {\bibfnamefont {M.}~\bibnamefont {Ricotti}},\ }\href
  {\doibase 10.1086/518998} {\bibfield  {journal} {\bibinfo  {journal}
  {Astrophys. J.}\ }\textbf {\bibinfo {volume} {665}},\ \bibinfo {pages} {1277}
  (\bibinfo {year} {2007})},\ \Eprint {http://arxiv.org/abs/astro-ph/0608642}
  {arXiv:astro-ph/0608642} \BibitemShut {NoStop}%
\bibitem [{\citenamefont {{Blecha}}\ and\ \citenamefont
  {{Loeb}}(2008)}]{Blecha:2008mg}%
  \BibitemOpen
  \bibfield  {author} {\bibinfo {author} {\bibfnamefont {L.}~\bibnamefont
  {{Blecha}}}\ and\ \bibinfo {author} {\bibfnamefont {A.}~\bibnamefont
  {{Loeb}}},\ }\href {\doibase 10.1111/j.1365-2966.2008.13790.x} {\bibfield
  {journal} {\bibinfo  {journal} {mnras}\ }\textbf {\bibinfo {volume} {390}},\
  \bibinfo {pages} {1311} (\bibinfo {year} {2008})},\ \Eprint
  {http://arxiv.org/abs/0805.1420} {arXiv:0805.1420} \BibitemShut {NoStop}%
\bibitem [{\citenamefont {Harada}\ and\ \citenamefont
  {Kimura}(2014)}]{Harada:2014vka}%
  \BibitemOpen
  \bibfield  {author} {\bibinfo {author} {\bibfnamefont {T.}~\bibnamefont
  {Harada}}\ and\ \bibinfo {author} {\bibfnamefont {M.}~\bibnamefont
  {Kimura}},\ }\href {\doibase 10.1088/0264-9381/31/24/243001} {\bibfield
  {journal} {\bibinfo  {journal} {Class. Quant. Grav.}\ }\textbf {\bibinfo
  {volume} {31}},\ \bibinfo {pages} {243001} (\bibinfo {year} {2014})},\
  \Eprint {http://arxiv.org/abs/1409.7502} {arXiv:1409.7502 [gr-qc]}
  \BibitemShut {NoStop}%
\bibitem [{\citenamefont {Sperhake}\ \emph {et~al.}(2008)\citenamefont
  {Sperhake}, \citenamefont {Cardoso}, \citenamefont {Pretorius}, \citenamefont
  {Berti},\ and\ \citenamefont {Gonzalez}}]{Sperhake:2008ga}%
  \BibitemOpen
  \bibfield  {author} {\bibinfo {author} {\bibfnamefont {U.}~\bibnamefont
  {Sperhake}}, \bibinfo {author} {\bibfnamefont {V.}~\bibnamefont {Cardoso}},
  \bibinfo {author} {\bibfnamefont {F.}~\bibnamefont {Pretorius}}, \bibinfo
  {author} {\bibfnamefont {E.}~\bibnamefont {Berti}}, \ and\ \bibinfo {author}
  {\bibfnamefont {J.~A.}\ \bibnamefont {Gonzalez}},\ }\href {\doibase
  10.1103/PhysRevLett.101.161101} {\bibfield  {journal} {\bibinfo  {journal}
  {Phys. Rev. Lett.}\ }\textbf {\bibinfo {volume} {101}},\ \bibinfo {pages}
  {161101} (\bibinfo {year} {2008})},\ \Eprint {http://arxiv.org/abs/0806.1738}
  {arXiv:0806.1738 [gr-qc]} \BibitemShut {NoStop}%
\bibitem [{\citenamefont {Sperhake}\ \emph {et~al.}(2013)\citenamefont
  {Sperhake}, \citenamefont {Berti}, \citenamefont {Cardoso},\ and\
  \citenamefont {Pretorius}}]{Sperhake:2012me}%
  \BibitemOpen
  \bibfield  {author} {\bibinfo {author} {\bibfnamefont {U.}~\bibnamefont
  {Sperhake}}, \bibinfo {author} {\bibfnamefont {E.}~\bibnamefont {Berti}},
  \bibinfo {author} {\bibfnamefont {V.}~\bibnamefont {Cardoso}}, \ and\
  \bibinfo {author} {\bibfnamefont {F.}~\bibnamefont {Pretorius}},\ }\href
  {\doibase 10.1103/PhysRevLett.111.041101} {\bibfield  {journal} {\bibinfo
  {journal} {Phys. Rev. Lett.}\ }\textbf {\bibinfo {volume} {111}},\ \bibinfo
  {pages} {041101} (\bibinfo {year} {2013})},\ \Eprint
  {http://arxiv.org/abs/1211.6114} {arXiv:1211.6114 [gr-qc]} \BibitemShut
  {NoStop}%
\bibitem [{\citenamefont {Healy}\ \emph {et~al.}(2016)\citenamefont {Healy},
  \citenamefont {Ruchlin}, \citenamefont {Lousto},\ and\ \citenamefont
  {Zlochower}}]{Healy:2015mla}%
  \BibitemOpen
  \bibfield  {author} {\bibinfo {author} {\bibfnamefont {J.}~\bibnamefont
  {Healy}}, \bibinfo {author} {\bibfnamefont {I.}~\bibnamefont {Ruchlin}},
  \bibinfo {author} {\bibfnamefont {C.~O.}\ \bibnamefont {Lousto}}, \ and\
  \bibinfo {author} {\bibfnamefont {Y.}~\bibnamefont {Zlochower}},\ }\href
  {\doibase 10.1103/PhysRevD.94.104020} {\bibfield  {journal} {\bibinfo
  {journal} {Phys. Rev.}\ }\textbf {\bibinfo {volume} {D94}},\ \bibinfo {pages}
  {104020} (\bibinfo {year} {2016})},\ \Eprint
  {http://arxiv.org/abs/1506.06153} {arXiv:1506.06153 [gr-qc]} \BibitemShut
  {NoStop}%
\bibitem [{\citenamefont {Ruchlin}\ \emph {et~al.}(2017)\citenamefont
  {Ruchlin}, \citenamefont {Healy}, \citenamefont {Lousto},\ and\ \citenamefont
  {Zlochower}}]{Ruchlin:2014zva}%
  \BibitemOpen
  \bibfield  {author} {\bibinfo {author} {\bibfnamefont {I.}~\bibnamefont
  {Ruchlin}}, \bibinfo {author} {\bibfnamefont {J.}~\bibnamefont {Healy}},
  \bibinfo {author} {\bibfnamefont {C.~O.}\ \bibnamefont {Lousto}}, \ and\
  \bibinfo {author} {\bibfnamefont {Y.}~\bibnamefont {Zlochower}},\ }\href
  {\doibase 10.1103/PhysRevD.95.024033} {\bibfield  {journal} {\bibinfo
  {journal} {Phys. Rev.}\ }\textbf {\bibinfo {volume} {D95}},\ \bibinfo {pages}
  {024033} (\bibinfo {year} {2017})},\ \Eprint {http://arxiv.org/abs/1410.8607}
  {arXiv:1410.8607 [gr-qc]} \BibitemShut {NoStop}%
\bibitem [{\citenamefont {Shibata}\ \emph {et~al.}(2008)\citenamefont
  {Shibata}, \citenamefont {Okawa},\ and\ \citenamefont
  {Yamamoto}}]{Shibata:2008rq}%
  \BibitemOpen
  \bibfield  {author} {\bibinfo {author} {\bibfnamefont {M.}~\bibnamefont
  {Shibata}}, \bibinfo {author} {\bibfnamefont {H.}~\bibnamefont {Okawa}}, \
  and\ \bibinfo {author} {\bibfnamefont {T.}~\bibnamefont {Yamamoto}},\ }\href
  {\doibase 10.1103/PhysRevD.78.101501} {\bibfield  {journal} {\bibinfo
  {journal} {Phys. Rev.}\ }\textbf {\bibinfo {volume} {D78}},\ \bibinfo {pages}
  {101501} (\bibinfo {year} {2008})},\ \Eprint {http://arxiv.org/abs/0810.4735}
  {arXiv:0810.4735 [gr-qc]} \BibitemShut {NoStop}%
\bibitem [{\citenamefont {Berti}\ \emph {et~al.}(2010)\citenamefont {Berti},
  \citenamefont {Cardoso}, \citenamefont {Hinderer}, \citenamefont {Lemos},
  \citenamefont {Pretorius} \emph {et~al.}}]{Berti:2010ce}%
  \BibitemOpen
  \bibfield  {author} {\bibinfo {author} {\bibfnamefont {E.}~\bibnamefont
  {Berti}}, \bibinfo {author} {\bibfnamefont {V.}~\bibnamefont {Cardoso}},
  \bibinfo {author} {\bibfnamefont {T.}~\bibnamefont {Hinderer}}, \bibinfo
  {author} {\bibfnamefont {M.}~\bibnamefont {Lemos}}, \bibinfo {author}
  {\bibfnamefont {F.}~\bibnamefont {Pretorius}},  \emph {et~al.},\ }\href
  {\doibase 10.1103/PhysRevD.81.104048} {\bibfield  {journal} {\bibinfo
  {journal} {Phys. Rev.}\ }\textbf {\bibinfo {volume} {D81}},\ \bibinfo {pages}
  {104048} (\bibinfo {year} {2010})},\ \Eprint {http://arxiv.org/abs/1003.0812}
  {arXiv:1003.0812 [gr-qc]} \BibitemShut {NoStop}%
\bibitem [{\citenamefont {Sperhake}\ \emph {et~al.}(2019)\citenamefont
  {Sperhake}, \citenamefont {Cook},\ and\ \citenamefont
  {Wang}}]{Sperhake:2019oaw}%
  \BibitemOpen
  \bibfield  {author} {\bibinfo {author} {\bibfnamefont {U.}~\bibnamefont
  {Sperhake}}, \bibinfo {author} {\bibfnamefont {W.}~\bibnamefont {Cook}}, \
  and\ \bibinfo {author} {\bibfnamefont {D.}~\bibnamefont {Wang}},\ }\href
  {\doibase 10.1103/PhysRevD.100.104046} {\bibfield  {journal} {\bibinfo
  {journal} {Phys. Rev. D}\ }\textbf {\bibinfo {volume} {100}},\ \bibinfo
  {pages} {104046} (\bibinfo {year} {2019})},\ \Eprint
  {http://arxiv.org/abs/1909.02997} {arXiv:1909.02997 [gr-qc]} \BibitemShut
  {NoStop}%
\bibitem [{\citenamefont {Bozzola}(2022)}]{Bozzola:2022uqu}%
  \BibitemOpen
  \bibfield  {author} {\bibinfo {author} {\bibfnamefont {G.}~\bibnamefont
  {Bozzola}},\ }\href {\doibase 10.1103/PhysRevLett.128.071101} {\bibfield
  {journal} {\bibinfo  {journal} {Phys. Rev. Lett.}\ }\textbf {\bibinfo
  {volume} {128}},\ \bibinfo {pages} {071101} (\bibinfo {year} {2022})},\
  \Eprint {http://arxiv.org/abs/2202.05310} {arXiv:2202.05310 [gr-qc]}
  \BibitemShut {NoStop}%
\bibitem [{\citenamefont {Page}(2023)}]{Page:2022bem}%
  \BibitemOpen
  \bibfield  {author} {\bibinfo {author} {\bibfnamefont {D.~N.}\ \bibnamefont
  {Page}},\ }\href {\doibase 10.1103/PhysRevD.107.064057} {\bibfield  {journal}
  {\bibinfo  {journal} {Phys. Rev. D}\ }\textbf {\bibinfo {volume} {107}},\
  \bibinfo {pages} {064057} (\bibinfo {year} {2023})},\ \Eprint
  {http://arxiv.org/abs/2212.03890} {arXiv:2212.03890 [gr-qc]} \BibitemShut
  {NoStop}%
\bibitem [{\citenamefont {Zlochower}\ \emph {et~al.}(2005)\citenamefont
  {Zlochower}, \citenamefont {Baker}, \citenamefont {Campanelli},\ and\
  \citenamefont {Lousto}}]{Zlochower:2005bj}%
  \BibitemOpen
  \bibfield  {author} {\bibinfo {author} {\bibfnamefont {Y.}~\bibnamefont
  {Zlochower}}, \bibinfo {author} {\bibfnamefont {J.~G.}\ \bibnamefont
  {Baker}}, \bibinfo {author} {\bibfnamefont {M.}~\bibnamefont {Campanelli}}, \
  and\ \bibinfo {author} {\bibfnamefont {C.~O.}\ \bibnamefont {Lousto}},\
  }\href {\doibase 10.1103/PhysRevD.72.024021} {\bibfield  {journal} {\bibinfo
  {journal} {Phys. Rev.}\ }\textbf {\bibinfo {volume} {D72}},\ \bibinfo {pages}
  {024021} (\bibinfo {year} {2005})},\ \Eprint
  {http://arxiv.org/abs/gr-qc/0505055} {arXiv:gr-qc/0505055} \BibitemShut
  {NoStop}%
\bibitem [{\citenamefont {Campanelli}\ \emph
  {et~al.}(2006{\natexlab{b}})\citenamefont {Campanelli}, \citenamefont
  {Lousto}, \citenamefont {Marronetti},\ and\ \citenamefont
  {Zlochower}}]{Campanelli:2005dd}%
  \BibitemOpen
  \bibfield  {author} {\bibinfo {author} {\bibfnamefont {M.}~\bibnamefont
  {Campanelli}}, \bibinfo {author} {\bibfnamefont {C.~O.}\ \bibnamefont
  {Lousto}}, \bibinfo {author} {\bibfnamefont {P.}~\bibnamefont {Marronetti}},
  \ and\ \bibinfo {author} {\bibfnamefont {Y.}~\bibnamefont {Zlochower}},\
  }\href@noop {} {\bibfield  {journal} {\bibinfo  {journal} {Phys. Rev. Lett.}\
  }\textbf {\bibinfo {volume} {96}},\ \bibinfo {pages} {111101} (\bibinfo
  {year} {2006}{\natexlab{b}})},\ \Eprint {http://arxiv.org/abs/gr-qc/0511048}
  {gr-qc/0511048} \BibitemShut {NoStop}%
\bibitem [{\citenamefont {Nakamura}\ \emph {et~al.}(1987)\citenamefont
  {Nakamura}, \citenamefont {Oohara},\ and\ \citenamefont
  {Kojima}}]{Nakamura87}%
  \BibitemOpen
  \bibfield  {author} {\bibinfo {author} {\bibfnamefont {T.}~\bibnamefont
  {Nakamura}}, \bibinfo {author} {\bibfnamefont {K.}~\bibnamefont {Oohara}}, \
  and\ \bibinfo {author} {\bibfnamefont {Y.}~\bibnamefont {Kojima}},\
  }\href@noop {} {\bibfield  {journal} {\bibinfo  {journal} {Prog. Theor. Phys.
  Suppl.}\ }\textbf {\bibinfo {volume} {90}},\ \bibinfo {pages} {1} (\bibinfo
  {year} {1987})}\BibitemShut {NoStop}%
\bibitem [{\citenamefont {Shibata}\ and\ \citenamefont
  {Nakamura}(1995)}]{Shibata95}%
  \BibitemOpen
  \bibfield  {author} {\bibinfo {author} {\bibfnamefont {M.}~\bibnamefont
  {Shibata}}\ and\ \bibinfo {author} {\bibfnamefont {T.}~\bibnamefont
  {Nakamura}},\ }\href@noop {} {\bibfield  {journal} {\bibinfo  {journal}
  {Phys. Rev.}\ }\textbf {\bibinfo {volume} {D52}},\ \bibinfo {pages} {5428}
  (\bibinfo {year} {1995})}\BibitemShut {NoStop}%
\bibitem [{\citenamefont {Baumgarte}\ and\ \citenamefont
  {Shapiro}(1998)}]{Baumgarte99}%
  \BibitemOpen
  \bibfield  {author} {\bibinfo {author} {\bibfnamefont {T.~W.}\ \bibnamefont
  {Baumgarte}}\ and\ \bibinfo {author} {\bibfnamefont {S.~L.}\ \bibnamefont
  {Shapiro}},\ }\href@noop {} {\bibfield  {journal} {\bibinfo  {journal} {Phys.
  Rev.}\ }\textbf {\bibinfo {volume} {D59}},\ \bibinfo {pages} {024007}
  (\bibinfo {year} {1998})},\ \Eprint {http://arxiv.org/abs/gr-qc/9810065}
  {gr-qc/9810065} \BibitemShut {NoStop}%
\bibitem [{cac()}]{cactus_web}%
  \BibitemOpen
  \href@noop {} {}\bibinfo {note} {Cactus Computational Toolkit home page: {\tt
  http://cactuscode.org}}\BibitemShut {NoStop}%
\bibitem [{\citenamefont {Schnetter}\ \emph {et~al.}(2004)\citenamefont
  {Schnetter}, \citenamefont {Hawley},\ and\ \citenamefont
  {Hawke}}]{Schnetter-etal-03b}%
  \BibitemOpen
  \bibfield  {author} {\bibinfo {author} {\bibfnamefont {E.}~\bibnamefont
  {Schnetter}}, \bibinfo {author} {\bibfnamefont {S.~H.}\ \bibnamefont
  {Hawley}}, \ and\ \bibinfo {author} {\bibfnamefont {I.}~\bibnamefont
  {Hawke}},\ }\href@noop {} {\bibfield  {journal} {\bibinfo  {journal} {Class.
  Quant. Grav.}\ }\textbf {\bibinfo {volume} {21}},\ \bibinfo {pages} {1465}
  (\bibinfo {year} {2004})},\ \Eprint {http://arxiv.org/abs/gr-qc/0310042}
  {gr-qc/0310042} \BibitemShut {NoStop}%
\bibitem [{\citenamefont {L{\"o}ffler}\ \emph {et~al.}(2012)\citenamefont
  {L{\"o}ffler}, \citenamefont {Faber}, \citenamefont {Bentivegna},
  \citenamefont {Bode}, \citenamefont {Diener}, \citenamefont {Haas},
  \citenamefont {Hinder}, \citenamefont {Mundim}, \citenamefont {Ott},
  \citenamefont {Schnetter}, \citenamefont {Allen}, \citenamefont
  {Campanelli},\ and\ \citenamefont {Laguna}}]{Loffler:2011ay}%
  \BibitemOpen
  \bibfield  {author} {\bibinfo {author} {\bibfnamefont {F.}~\bibnamefont
  {L{\"o}ffler}}, \bibinfo {author} {\bibfnamefont {J.}~\bibnamefont {Faber}},
  \bibinfo {author} {\bibfnamefont {E.}~\bibnamefont {Bentivegna}}, \bibinfo
  {author} {\bibfnamefont {T.}~\bibnamefont {Bode}}, \bibinfo {author}
  {\bibfnamefont {P.}~\bibnamefont {Diener}}, \bibinfo {author} {\bibfnamefont
  {R.}~\bibnamefont {Haas}}, \bibinfo {author} {\bibfnamefont {I.}~\bibnamefont
  {Hinder}}, \bibinfo {author} {\bibfnamefont {B.~C.}\ \bibnamefont {Mundim}},
  \bibinfo {author} {\bibfnamefont {C.~D.}\ \bibnamefont {Ott}}, \bibinfo
  {author} {\bibfnamefont {E.}~\bibnamefont {Schnetter}}, \bibinfo {author}
  {\bibfnamefont {G.}~\bibnamefont {Allen}}, \bibinfo {author} {\bibfnamefont
  {M.}~\bibnamefont {Campanelli}}, \ and\ \bibinfo {author} {\bibfnamefont
  {P.}~\bibnamefont {Laguna}},\ }\href@noop {} {\bibfield  {journal} {\bibinfo
  {journal} {Class. Quant. Grav.}\ }\textbf {\bibinfo {volume} {29}},\ \bibinfo
  {pages} {115001} (\bibinfo {year} {2012})},\ \Eprint
  {http://arxiv.org/abs/1111.3344} {arXiv:1111.3344 [gr-qc]} \BibitemShut
  {NoStop}%
\bibitem [{ein()}]{einsteintoolkit}%
  \BibitemOpen
  \href@noop {} {}\bibinfo {note} {Einstein Toolkit home page: {\tt
  http://einsteintoolkit.org}}\BibitemShut {NoStop}%
\bibitem [{\citenamefont {Ansorg}\ \emph {et~al.}(2004)\citenamefont {Ansorg},
  \citenamefont {Br\"ugmann},\ and\ \citenamefont {Tichy}}]{Ansorg:2004ds}%
  \BibitemOpen
  \bibfield  {author} {\bibinfo {author} {\bibfnamefont {M.}~\bibnamefont
  {Ansorg}}, \bibinfo {author} {\bibfnamefont {B.}~\bibnamefont {Br\"ugmann}},
  \ and\ \bibinfo {author} {\bibfnamefont {W.}~\bibnamefont {Tichy}},\
  }\href@noop {} {\bibfield  {journal} {\bibinfo  {journal} {Phys. Rev.}\
  }\textbf {\bibinfo {volume} {D70}},\ \bibinfo {pages} {064011} (\bibinfo
  {year} {2004})},\ \Eprint {http://arxiv.org/abs/gr-qc/0404056}
  {gr-qc/0404056} \BibitemShut {NoStop}%
\bibitem [{\citenamefont {Thornburg}(2004)}]{Thornburg2003:AH-finding}%
  \BibitemOpen
  \bibfield  {author} {\bibinfo {author} {\bibfnamefont {J.}~\bibnamefont
  {Thornburg}},\ }\href {\doibase 10.1088/0264-9381/21/2/026} {\bibfield
  {journal} {\bibinfo  {journal} {Class. Quant. Grav.}\ }\textbf {\bibinfo
  {volume} {21}},\ \bibinfo {pages} {743} (\bibinfo {year} {2004})},\ \Eprint
  {http://arxiv.org/abs/gr-qc/0306056} {gr-qc/0306056} \BibitemShut {NoStop}%
\bibitem [{\citenamefont {Campanelli}\ \emph
  {et~al.}(2007{\natexlab{c}})\citenamefont {Campanelli}, \citenamefont
  {Lousto}, \citenamefont {Zlochower}, \citenamefont {Krishnan},\ and\
  \citenamefont {Merritt}}]{Campanelli:2006fy}%
  \BibitemOpen
  \bibfield  {author} {\bibinfo {author} {\bibfnamefont {M.}~\bibnamefont
  {Campanelli}}, \bibinfo {author} {\bibfnamefont {C.~O.}\ \bibnamefont
  {Lousto}}, \bibinfo {author} {\bibfnamefont {Y.}~\bibnamefont {Zlochower}},
  \bibinfo {author} {\bibfnamefont {B.}~\bibnamefont {Krishnan}}, \ and\
  \bibinfo {author} {\bibfnamefont {D.}~\bibnamefont {Merritt}},\ }\href@noop
  {} {\bibfield  {journal} {\bibinfo  {journal} {Phys. Rev.}\ }\textbf
  {\bibinfo {volume} {D75}},\ \bibinfo {pages} {064030} (\bibinfo {year}
  {2007}{\natexlab{c}})},\ \Eprint {http://arxiv.org/abs/gr-qc/0612076}
  {gr-qc/0612076} \BibitemShut {NoStop}%
\bibitem [{\citenamefont {Campanelli}\ and\ \citenamefont
  {Lousto}(1999)}]{Campanelli:1998jv}%
  \BibitemOpen
  \bibfield  {author} {\bibinfo {author} {\bibfnamefont {M.}~\bibnamefont
  {Campanelli}}\ and\ \bibinfo {author} {\bibfnamefont {C.~O.}\ \bibnamefont
  {Lousto}},\ }\href {\doibase 10.1103/PhysRevD.59.124022} {\bibfield
  {journal} {\bibinfo  {journal} {Phys. Rev.}\ }\textbf {\bibinfo {volume}
  {D59}},\ \bibinfo {pages} {124022} (\bibinfo {year} {1999})},\ \Eprint
  {http://arxiv.org/abs/gr-qc/9811019} {arXiv:gr-qc/9811019 [gr-qc]}
  \BibitemShut {NoStop}%
\bibitem [{\citenamefont {Lousto}\ and\ \citenamefont
  {Zlochower}(2007)}]{Lousto:2007mh}%
  \BibitemOpen
  \bibfield  {author} {\bibinfo {author} {\bibfnamefont {C.~O.}\ \bibnamefont
  {Lousto}}\ and\ \bibinfo {author} {\bibfnamefont {Y.}~\bibnamefont
  {Zlochower}},\ }\href@noop {} {\bibfield  {journal} {\bibinfo  {journal}
  {Phys. Rev.}\ }\textbf {\bibinfo {volume} {D76}},\ \bibinfo {pages}
  {041502(R)} (\bibinfo {year} {2007})},\ \Eprint
  {http://arxiv.org/abs/gr-qc/0703061} {gr-qc/0703061} \BibitemShut {NoStop}%
\bibitem [{\citenamefont {Nakano}\ \emph {et~al.}(2015)\citenamefont {Nakano},
  \citenamefont {Healy}, \citenamefont {Lousto},\ and\ \citenamefont
  {Zlochower}}]{Nakano:2015pta}%
  \BibitemOpen
  \bibfield  {author} {\bibinfo {author} {\bibfnamefont {H.}~\bibnamefont
  {Nakano}}, \bibinfo {author} {\bibfnamefont {J.}~\bibnamefont {Healy}},
  \bibinfo {author} {\bibfnamefont {C.~O.}\ \bibnamefont {Lousto}}, \ and\
  \bibinfo {author} {\bibfnamefont {Y.}~\bibnamefont {Zlochower}},\ }\href
  {\doibase 10.1103/PhysRevD.91.104022} {\bibfield  {journal} {\bibinfo
  {journal} {Phys. Rev.}\ }\textbf {\bibinfo {volume} {D91}},\ \bibinfo {pages}
  {104022} (\bibinfo {year} {2015})},\ \Eprint
  {http://arxiv.org/abs/1503.00718} {arXiv:1503.00718 [gr-qc]} \BibitemShut
  {NoStop}%
\bibitem [{\citenamefont {Healy}\ \emph {et~al.}(2009)\citenamefont {Healy},
  \citenamefont {Herrmann}, \citenamefont {Hinder}, \citenamefont {Shoemaker},
  \citenamefont {Laguna},\ and\ \citenamefont {Matzner}}]{Healy:2008js}%
  \BibitemOpen
  \bibfield  {author} {\bibinfo {author} {\bibfnamefont {J.}~\bibnamefont
  {Healy}}, \bibinfo {author} {\bibfnamefont {F.}~\bibnamefont {Herrmann}},
  \bibinfo {author} {\bibfnamefont {I.}~\bibnamefont {Hinder}}, \bibinfo
  {author} {\bibfnamefont {D.~M.}\ \bibnamefont {Shoemaker}}, \bibinfo {author}
  {\bibfnamefont {P.}~\bibnamefont {Laguna}}, \ and\ \bibinfo {author}
  {\bibfnamefont {R.~A.}\ \bibnamefont {Matzner}},\ }\href {\doibase
  10.1103/PhysRevLett.102.041101} {\bibfield  {journal} {\bibinfo  {journal}
  {Phys. Rev. Lett.}\ }\textbf {\bibinfo {volume} {102}},\ \bibinfo {pages}
  {041101} (\bibinfo {year} {2009})},\ \Eprint {http://arxiv.org/abs/0807.3292}
  {arXiv:0807.3292 [gr-qc]} \BibitemShut {NoStop}%
\bibitem [{\citenamefont {Sperhake}\ \emph {et~al.}(2011)\citenamefont
  {Sperhake}, \citenamefont {Berti}, \citenamefont {Cardoso}, \citenamefont
  {Pretorius},\ and\ \citenamefont {Yunes}}]{Sperhake:2010uv}%
  \BibitemOpen
  \bibfield  {author} {\bibinfo {author} {\bibfnamefont {U.}~\bibnamefont
  {Sperhake}}, \bibinfo {author} {\bibfnamefont {E.}~\bibnamefont {Berti}},
  \bibinfo {author} {\bibfnamefont {V.}~\bibnamefont {Cardoso}}, \bibinfo
  {author} {\bibfnamefont {F.}~\bibnamefont {Pretorius}}, \ and\ \bibinfo
  {author} {\bibfnamefont {N.}~\bibnamefont {Yunes}},\ }\href {\doibase
  10.1103/PhysRevD.83.024037} {\bibfield  {journal} {\bibinfo  {journal} {Phys.
  Rev.}\ }\textbf {\bibinfo {volume} {D83}},\ \bibinfo {pages} {024037}
  (\bibinfo {year} {2011})},\ \Eprint {http://arxiv.org/abs/1011.3281}
  {arXiv:1011.3281 [gr-qc]} \BibitemShut {NoStop}%
\bibitem [{\citenamefont {Lousto}\ and\ \citenamefont
  {Zlochower}(2011{\natexlab{b}})}]{Lousto:2010xk}%
  \BibitemOpen
  \bibfield  {author} {\bibinfo {author} {\bibfnamefont {C.~O.}\ \bibnamefont
  {Lousto}}\ and\ \bibinfo {author} {\bibfnamefont {Y.}~\bibnamefont
  {Zlochower}},\ }\href {\doibase 10.1103/PhysRevD.83.024003} {\bibfield
  {journal} {\bibinfo  {journal} {Phys. Rev.}\ }\textbf {\bibinfo {volume}
  {D83}},\ \bibinfo {pages} {024003} (\bibinfo {year} {2011}{\natexlab{b}})},\
  \Eprint {http://arxiv.org/abs/1011.0593} {arXiv:1011.0593 [gr-qc]}
  \BibitemShut {NoStop}%
\bibitem [{\citenamefont {Ciarfella}\ \emph {et~al.}()\citenamefont
  {Ciarfella}, \citenamefont {Healy},\ and\ \citenamefont
  {Lousto}}]{CHLinprep}%
  \BibitemOpen
  \bibfield  {author} {\bibinfo {author} {\bibfnamefont {A.}~\bibnamefont
  {Ciarfella}}, \bibinfo {author} {\bibfnamefont {J.}~\bibnamefont {Healy}}, \
  and\ \bibinfo {author} {\bibfnamefont {C.~O.}\ \bibnamefont {Lousto}},\
  }\href@noop {} {}\bibinfo {note} {In preparation}\BibitemShut {NoStop}%
\end{thebibliography}%

\end{document}